\pdfoutput=1
\documentclass[%
 reprint,
superscriptaddress,
 amsmath,amssymb,
 prl
]{revtex4-1}

\usepackage{graphicx}% Include figure files
\usepackage{dcolumn}% Align table columns on decimal point
\usepackage{bm}% bold math
\usepackage{hyperref}% add hypertext capabilities
\usepackage{newtxtext}
\usepackage{xcolor}
\usepackage{braket}
\usepackage{tikz}
\usetikzlibrary{shapes.misc}
\usetikzlibrary{decorations.pathmorphing}
\tikzset{snake it/.style={decorate, decoration=snake}}
\tikzset{cross/.style={cross out, draw=black, minimum size=2*(#1-\pgflinewidth), inner sep=0pt, outer sep=0pt},
cross/.default={1pt}}

\let\oldparagraph\paragraph
\renewcommand{\paragraph}[1]{\oldparagraph{#1---}\hspace{-1em}}
\usepackage{subcaption}

\begin{document}

\title{Replica Instantons from Axion-like Coupling}

\author{Kantaro Ohmori}
 \email{komori@scgp.stonybrook.edu}
\affiliation{%
Simons Center for Geometry and Physics, SUNY
}%
\date{\today}% It is always \today, today,
             %  but any date may be explicitly specified

\begin{abstract}
   We find a phenomenon in a non-gravitational gauge theory analogous to the replica wormhole in a quantum gravity theory.
   We consider a reservoir of a scalar field coupled with a gauge theory contained in a region with boundary by an axion-like coupling.
   When the replica trick is used to compute the entanglement entropy for a subregion in the reservoir,
   a tuple of instantons distributed across the replica sheets gives a non-perturbative contribution.
   As an explicit and solvable example, we consider a discrete scalar field coupled to a 2d pure gauge theory and observe how the replica instantons reproduce the entropy directly calculated from the reduced density matrix.
   In addition, we notice that the entanglement entropy can detect the confinement of a 2d gauge theory.
\end{abstract}

\maketitle

%\tableofcontents

\paragraph{Introduction}
One of the most significant recent developments in the study of quantum gravity is the island conjecture \cite{Penington:2019npb,Almheiri:2019psf} and the replica wormhole \cite{Penington:2019kki,Almheiri:2019qdq} that reproduces the conjecture.
These were used to derive the Page curve of the entropy of the radiation from a black hole, and are considered to be a key milestone to resolve the information paradox.

%While these achievements are big steps forward, it is not easy to contract a tractable UV completed example, especially in dimensions higher than 2.
%One possible strategy is to resort to non-gravitational theory and find some analog of the phenomena.
%As the UV completion of a field theory is more controllable than that of a gravity theory, one might obtain a better understanding of a part of how the replica 
In this letter, we find an analog in a non-gravitational gauge theory for the replica wormhole. 
This might indicate that some part of the black hole evaporation process has an analogy in gauge theory and can be studied in it.
In particular, we perform the exact computation for 2d pure gauge theories. There, we observe that the entanglement we consider can detect the confinement of the theory.

The state $\ket{\psi}$ considered in this letter is artificial, without a clear experimental meaning. Finding a more physical setup, preferably with some connection or analog to the black hole physics, is remained for future studies.

The replica wormhole is a wormhole connecting different sheets of the replicas in the replica trick computation of entanglement entropy.
Such a configuration should be considered if the gravitational path-integral involves the summation over the topology of the geometry.
In a non-gravitational theory, instead of the topology of the geometry, the topology of the gauge group bundle can be nontrivial on the replica manifold.

The configuration of interest is that an instanton is present on one of the replica sheets, at the same time an anti-instanton appears on another. Under the presence of an axion-like coupling, the total instanton number on the whole replica manifold is constrained to be zero, so that such a pair, or in general a tuple, of instantons causes effects not local to any one of sheets. 
Therefore, we regard this non-local distribution of instantons as an analog of the replica wormhole.
The analogy between a wormhole and a pair of instantons is also noticed in \cite{Yonekura:2020ino}.

\paragraph{General setup}
Consider a scalar field $\phi$ coupled to a $\mathbb{Z}$-valued topological number density $c(A)$ of a gauge field $A$ on a spacetime $X$:
\begin{equation}
    \mathrm{i}\int_{X} \phi \, c(A).
    \label{eq:axion}
\end{equation}
For example, $c(A)$ can be the instanton density $\mathrm{Tr}F\wedge F$ if the spacetime dimension is four, and the coupling is that of the axion.\footnote{We can consider a sigma model that admits a theta term instead of a gauge theory. The general argument of this letter applies to that case.}
We assume that the manifold $X$ has a boundary and is embedded into a manifold $Y$ of the same dimensions.
While we make $\phi$ live over the whole manifold $Y$, the gauge field $A$ (and matter fields charged under the gauge group) is contained inside the submanifold $X$.
We impose the Dirichlet boundary condition on $A$ on $\partial X$, i.e.\ $A|_{\partial X} = 0$ .
We regard the scalar $\phi$ outside of $X$ as a reservoir, and consider the entropy of a subregion of the reservoir.
This configuration mimics the setup used in the gravity theories.

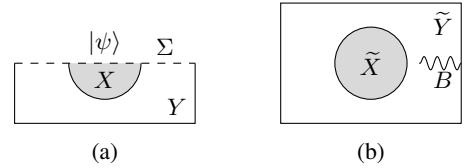
\begin{figure}[t]
    \centering
    \begin{subfigure}[b]{.5\linewidth}
    \begin{tikzpicture}[scale = .8,baseline=-40]
        \draw (3,-1) -- ++(0,-1) -- ++(-3,0) --++(0,1);
        \draw[fill = gray!30] (2.1,-1) coordinate (c) arc (0:-180:.6cm);
        %\draw[snake it,segment length=5] (0,-1) -- ++ (.7,0);
        %\draw[snake it,segment length=5] (3,-1) -- ++ (-.7,0);
        \draw[dashed] (0,-1) --++ (3,0);
        \node[anchor = south east] at (3,-2) {$Y$};
        \node[anchor = north] at (1.5,-1) {$X$};
        \node[anchor = south] at (1.5,-1) {$\ket{\psi}$};
        \node[anchor = south] at (2.5,-1) {$\Sigma$};
    \end{tikzpicture}
    \caption{}
    \label{fig:psi}
    \end{subfigure}
    \begin{subfigure}[b]{.3\linewidth}
    \begin{tikzpicture}[scale = .8]
        \draw (0,0) -- ++ (3,0) -- ++(0,-2) -- ++(-3,0) -- cycle;
        \draw[fill = gray!30] (1.5,-1) coordinate (c) circle (.6cm);
        %\draw[snake it,segment length=5] (0,-1) -- ++ (.7,0);
        \draw[snake it,segment length=5] (3,-1) -- ++ (-.7,0);
        \node[anchor = north east] at (3,0) {$\widetilde{Y}$};
        \node at (c) {$\widetilde{X}$};
        \node[anchor = north] at (2.7,-1) {$B$};
    \end{tikzpicture}
    \caption{}
    \label{fig:replica}
    \end{subfigure}
    \caption{(a) A spacetime geometry defining the pure state $\psi$. The shaded region is where the gauge theory lives and the white region is the reservoir of scalar field.  (b) A single sheet of the replica manifold $R_n$.}
\end{figure}

To prepare a state, we let $Y$ has the spatial boundary $\Sigma$ that intersects with $X$, and we consider the state $\ket{\psi}$ defined by the Euclidean path integral over $Y$ (See Fig.\ \ref{fig:psi}).\footnote{
    The geometry might admit a quench picture; the gauge field $A$ is frozen due an additional Higgs field with a large potential, but in the region $X$ a huge positive mass-squared is added to the Higgs field.
}
We take a subregion $B$ of $\Sigma$ outside of $X$, and consider the entanglement entropy of the reduced density matrix $\rho_B = \mathrm{Tr}_{B^c}\ket{\psi}\bra{\psi}$, where $B^c$ is the complement of $B$ in $\Sigma$.

The convenient way of computing the entanglement entropy is the replica trick. We let $\widetilde{Y}$ denotes the double $Y \cup \overline{Y}$, where $\overline{Y}$ is the orientation reversal of $Y$ and the two manifolds are glued along the spatial boundary $\Sigma$.
A replica sheet is $\tilde{Y}$ with the cut along the subregion $B$ of $\Sigma$ (See Fig.\ \ref{fig:replica}).
$\widetilde{X}$ in the figure is the double $X\cup \overline{X}$ of $X$ glued along $X\cap \Sigma$, and is a codimension-0 submanifold of $\overline{Y}$.
The $n$-th replica manifold $R_n$ is constructed by glueing $n$ copies of the replica sheet along the cut $B$.
Then, the trace of the $n$-th power of $\rho_B$ is expressed by the partition functions:
\begin{equation}
    \mathrm{Tr}\rho_B^n = \frac{Z[R_n]}{Z[R_1]^n},
    \label{eq:replicaRenyi}
\end{equation}
with which the R\'enyi entropy $S^{(n)}_B$ is defined by
\begin{equation}
    S^{(n)} = \frac{1}{1-n}\mathrm{log}\mathrm{Tr}\rho_B^n.
\end{equation}
The von Neumann entropy $S^{\text{vN}}_B$ is the limit $\lim_{n\to 1}S^{(n)}_B$.

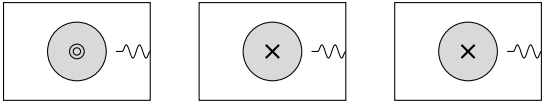
\begin{figure}[t]
    \centering
    \begin{tikzpicture}[scale = .65]
        \draw (0,0) coordinate(a) -- ++ (3,0) -- ++(0,-2) -- ++(-3,0) -- cycle;
        \draw[fill = gray!30] (a) ++ (1.5,-1)  circle (.6cm);
        %\draw[snake it,segment length=5] (0,-1) -- ++ (.7,0);
        \draw[snake it,segment length=5] (a) ++ (3,-1) -- ++ (-.7,0);
        \draw (a) ++ (1.5,-1) node[cross,thick,minimum size = 5pt] {};
        \draw (4,0) coordinate(b) -- ++ (3,0) -- ++(0,-2) -- ++(-3,0) -- cycle;
        \draw[fill = gray!30] (b) ++ (1.5,-1)  circle (.6cm);
        %\draw[snake it,segment length=5] (0,-1) -- ++ (.7,0);
        \draw[snake it,segment length=5] (b) ++ (3,-1) -- ++ (-.7,0);
        \draw (b) ++ (1.5,-1) node[cross,thick,minimum size = 5pt] {};
        \draw (-4,0) coordinate(c) -- ++ (3,0) -- ++(0,-2) -- ++(-3,0) -- cycle;
        \draw[fill = gray!30] (c) ++ (1.5,-1)  circle (.6cm);
        %\draw[snake it,segment length=5] (0,-1) -- ++ (.7,0);
        \draw[snake it,segment length=5] (c) ++ (3,-1) -- ++ (-.7,0);
        \draw (c) ++ (1.5,-1) circle (.075);
        \draw (c) ++ (1.5,-1) circle (.15);
    \end{tikzpicture}
    
    \caption{A replica instanton configuration for $n=3$. The double circle in the leftmost sheet denotes an instanton with instanton number 2, and each of the crosses in the middle and the rightmost sheets is a unit anti-instanton.}
    \label{fig:r_inst}
\end{figure}

If the scalar $\phi$ has the shift symmetry broken only by the interaction \eqref{eq:axion},  the topological number $\int_X c(A)$ is effectively constrained to be zero, since the integration over the constant mode of $\phi$ gives the delta function $\delta(\int_X c(A))$.
On a replica manifold $R_n$, however, the constant come of $\phi$ is constant over  the whole of $R_n$, and other modes are suppressed by the kinetic term.
This constant mode only constrains the \emph{total sum} of topological numbers: 
\begin{equation}
    \sum_{i = 1}^n k_i= 0, 
    \label{eq:k_const}
\end{equation}
where $k_i = \int_{\widetilde{X}_i} c(A)$ is the topological number on the $i$-th copy of $\widetilde{X}$. 
We write the contribution to $Z[R_n]$ from the configurations with topological numbers $(k_1,\cdots, k_n)$ by $Z[R_n,(k_1,\cdots,k_n)]$, so that 
\begin{equation}
    Z[R_n] = \sum_{\substack{k_1,\cdots,k_n\\ \sum_i k_i =0}}Z[R_n,(k_1,\cdots,k_n)].
    \label{eq:ZRn}
\end{equation}
The term other than $R_n^{(0,\cdots,0)}$ has a nontrivial topology of the bundles and give non-perturbative corrections. 
We call such a configuration ``replica instantons".
See Fig.\ \ref{fig:r_inst}) for an example.
Note that because of the constraint \eqref{eq:k_const}, the non-perturbative gauge theory contribution does not factorize into the product of the contributions from each copy $\widetilde{X}_i$, and thus it is not localized in any one of the copies.

\paragraph{Discrete scalar reservoir}
So far we have been assuming that the scalar $\phi$ takes continuous values. To compute the replica partition function, we have to do the path-integral of the scalar field on the replica manifold, which is in general not easy.
To simplify the computation, in this letter we instead make the scalar field takes the discrete values: $\phi = 0, \frac{2\pi}{q},\cdots, \frac{(q-1)2\pi}{q}$ $(\!\!\!\!\mod 2\pi)$ for some $q$. In other words, $\phi$ is a $\mathbb{Z}_q$-valued field.
This can be achieved either by imposing a potential $\cos(q\phi)$ with a big coefficient to a periodic scalar $\phi$, or by considering the lagrangian $q\phi F'$ and integrating the auxiliary gauge field $F'$(not to be confused with the gauge field on $X$) out.
Such a discrete scalar field defines a topological field theory, and 
the Hilbert space of the topological theory has dimension $q$ and is spanned by the coherent states $\ket{\ell}$, $\ell = 0,\cdots,q-1$ with $e^{\mathrm{i} \phi}\ket{\ell} = e^{2\pi \mathrm{i} \ell/q}\ket{\ell}$.

With the discrete scalar $\phi$, the path-integral over the replica manifold is replaced by a finite sum over $\mathbb{Z}_q$. We assume that the boundary condition of $\phi$ on the boundary of $B$ (and at the infinity of $\tilde{M}$ if it is non-compact) is Neumann so that the summation over the $\mathbb{Z}_q$ remains.\footnote{
    The boundary condition at the edge of the subregion in the replica manifold is regarded as a choice of UV regularization of the entanglement entropy \cite{Ohmori:2014eia}. In the context of topological field theory, the same boundary condition was used in \cite{Donnelly:2018ppr,Donnelly:2019zde}.
}
Another advantage to use the $\mathbb{Z}_q$ valued $\phi$ is that we can use a topological number defined only modulo $q$ as $c(A)$ in the coupling \eqref{eq:axion}. Accordingly, the ``path-sum" of $\phi$ imposes the constraint \eqref{eq:k_const} modulo $q$.

As the field $\phi$ now does not fluctuate, the evaluation of the partition function $Z[R_n]$ reduces to the partition function of the gauge theory:
\begin{equation}
    Z[R_n,(k_1,\cdots,k_n)] = q\prod_i Z_\text{gauge} [\tilde{X},k_i]
    \label{eq:ZRnk}
\end{equation}
where $Z_\text{gauge}[\tilde{X},k]$ denotes the partition function of the gauge theory on $\widetilde{X}$ in which only the bundles with $\int_{\widetilde{X}}c(A) = k \text{ mod $q$}$ are summed over. The factor $q$ comes from the summation over the values of $\phi$.
Substituting \eqref{eq:ZRnk} into \eqref{eq:ZRn},
we get 
\begin{equation}
    Z[R_n] = \sum_{\ell=0}^{q-1}(Z_\text{gauge}^{(\ell)}[\widetilde{X}])^n,
\end{equation}
where $Z_\text{gauge}^{(\ell)}[\widetilde{X}]$ is the discrete Fourier transform of $Z_\text{gauge}[\widetilde{X},k]$:
\begin{equation}
    Z_\text{gauge}^{(\ell)}[\widetilde{X}] = \sum_k Z_\text{gauge}[\widetilde{X},k] e^{2\pi\mathrm{i}k \ell/q}.
    \label{eq:Zell}
\end{equation}
This is also the partition function of the gauge theory with topological interaction $\frac{2\pi \ell}{q} \int c(A)$ added.
Thus the trace of $n$-th power of the reduced density matrix is 
\begin{equation}
    \mathrm{Tr}\rho_B^n = \frac{\sum_\ell \left(Z_\text{gauge}^{(\ell)}[\widetilde{X}]\right)^n}{\left(\sum_\ell Z_\text{gauge}^{(\ell)}[\widetilde{X}]\right)^n}.
    \label{eq:rhoB1}
\end{equation}

\paragraph{Direct computation of $\rho_B$}
The result \eqref{eq:rhoB1} suggests the simple form of the density matrix:
\begin{equation}
    \rho_B = \frac1{\sum_\ell Z_\text{gauge}^{(\ell)}[\widetilde{X}]}\mathrm{diag}(Z_\text{gauge}^{(0)}[\widetilde{X}], \cdots, Z_\text{gauge}^{(q-1)}[\widetilde{X}]).
    \label{eq:rhoB2}
\end{equation}
We can directly obtain this density matrix.
For a fixed value of $\phi = \ell$, the path-integral of the gauge theory over the manifold $X$ in Fig.\ \ref{fig:psi} defines a state $\ket{X}_\ell$ whose norm is $Z^{(\ell)}_\text{gauge}[\widetilde{X}]$.
The state $\ket{\psi}$ is 
\begin{equation}
    \ket{\psi} = \sum_\ell \ket{X}_\ell \otimes \ket{\ell}.
\end{equation} 
To define the reduced density matrix, we have to define a map \cite{Ohmori:2014eia}
\begin{equation}
    f: \mathcal{H}_\Sigma \to \mathcal{H}_{B^c}\otimes \mathcal{H}_B.
\end{equation}
In the replica computation we used the Neumann boundary condition at the boundary of $B$.
With the boundary condition, the Hilbert space $\mathcal{H}_B$ is spanned by the coherent states $\ket{\ell}_B$. 
Likewise, the Hilbert space $\mathcal{H}^\phi_{B^c}$ of the region $B^c$ prejected to the $\phi$-sector is also spanned by the coherent states $\ket{\ell}_{B^c}$.
 Then, the map $f$ is
\begin{equation}
    f:\ket{\ell} \mapsto \ket{\ell}_{B^c}\otimes \ket{\ell}_B.
\end{equation}
Therefore, the reduced density matrix is
\begin{equation}
    \begin{split}
    \rho_B &= \frac1{\braket{\psi|f^\dag f|\psi}}\mathrm{Tr}_{\mathcal{H}_{B^c}}f\ket{\psi}\bra{\psi}f^\dag\\
    %&= \mathrm{Tr}_{\mathcal{H}_{B^c}}(\sum_\ell \ket{X}_\ell\otimes\ket{\ell}_{B^c}\otimes \ket{\ell}_B)(\sum_\ell' \ket{X}_{\ell'}\otimes\ket{\ell'}_{B^c}\otimes \ket{\ell'}_B)\\
    &= \frac1{\sum_{\ell'} Z_\text{gauge}^{(\ell')}[\widetilde{X}]}\sum_\ell Z_\text{gauge}^{(\ell)}[\widetilde{X}] \ket{\ell}_B\bra{\ell}_B.
    \end{split}
    \label{eq:rhoB3}
\end{equation}
which is \eqref{eq:rhoB2}.

\paragraph{2d abelian gauge field}
First, we take the gauge group to be $U(1)$ and set the topological number to be the monopole number: $c(A) = F$.
We follow the analysis in \cite{Komargodski:2020mxz}.
For a fixed scalar field $\phi = \frac{2\pi \ell}{q}$, the theory is the $U(1)$ gauge theory with the theta angle $\theta =\frac{2\pi \ell}{q}$.

If the theory is put on an interval $I$ with Dirichlet boundary condition and the temporal gauge $A_0 = 0$ is imposed, the theory reduced to the quantum mechanics of the holonomy $G(t) = \int_I A$ with the action
\begin{equation}
    2\pi L \int \mathrm{d}t \left(\frac{1}{2g^2}\dot{G}^2 +\frac{\theta}{2\pi} \dot{G}\right),
\end{equation}
where $L$ is the length of the interval and $g$ is the coupling constant.
The Hamiltonian and the energy levels are 
\begin{gather}
    H = \frac12 \frac{g^2}{2\pi L}(\Pi_G - \theta L)^2,\label{eq:H}\\
    E_j = \frac12\frac{g^2 L}{2\pi}(2\pi j -\theta)^2,
\end{gather}
where $\Pi_G$ is the canonical momentum for $G$ and  $j\in\mathbb{Z}$. We write the corresponding normalized eigenstate by $\ket{j}$ which also diagonalize $\Pi_G$: $\Pi_G \ket{j} = g^2(j-\frac{\theta}{2\pi})$.
We take the the manifold $X$ in Fig.\ \ref{fig:psi} to be a rectangular, $X = [-t_0,0]\times [0,L]$ where the first factor is regarded as the time direction and the state $\ket{\psi}$ is prepared at $t=0$. 
As we impose the Dirichlet boundary condition at $\{-t_0\}\times [0,L]$, the holonomy $G(-t_0)$ is zero.
As the energy eigenstate $\ket{j}$ diagonalizes $\Pi_G$, 
the state at $t= -t_0$, which has eigenvalue zero of $G$, is
\begin{equation}
    \ket{t= -t_0} = \sum_j \ket{j}.
\end{equation}
Thus, the space time $X$ and the theta angle $\theta = \frac{2\pi \ell}{q}$ prepares the state
\begin{equation}
    \ket{X}_\ell = \sum_j e^{-a'(j-\frac{\ell}{q})^2} \ket{j},
\end{equation}
where $a' = \frac12 g^2 L t_0$ is the dimensionless combination of the area of $X$ and the coupling $g$.
As a 2d pure gauge theory is invariant under area-preserving diffeomorphisms, this states does not depends on the shape of $X$ other than the area.
The partition function of the double $\widetilde{X}$ is 
\begin{equation}
    Z_{U(1)}^{(\ell)}[\tilde{X}] = \sum_{j\in \mathbb{Z}} e^{-a (j-\frac{\ell}{q})^2},
\end{equation}
where $a = 2a'$ is the dimensionless area of $\tilde{X}$.

The contribution $Z_{U(1)}[\tilde{X},k]$ form each topological class of the bundles can be obtained by reversing the discrete Fourier transform \eqref{eq:Zell}:
\begin{equation}
    Z_{U(1)}[\tilde{X},k] = \frac1q\sum_{j\in \frac1q \mathbb{Z}} e^{-a j^2 + 2\pi \mathrm{i} jk} = \frac1q\vartheta(\frac kq,\frac{\mathrm{i}a}{\pi q^2}),
    \label{eq:ZXk}
\end{equation}
where $\vartheta(z,\tau) = \sum_{\tilde{\jmath}\in \mathbb{Z}} e^{\pi \mathrm{i}(\tau \tilde{\jmath}^2+2 z \tilde{\jmath})}$ is the elliptic theta function.
In particular, $Z_{U(1)}[\tilde{X},0]$ is the partition function of the $U(1)$ gauge theory with the coupling $g/q$ and $\theta = 0 $. In other words, the gauge group of $Z_{U(1)}[\tilde{X},0]$ is the extension
\begin{equation}
    1 \to \mathbb{Z}_q \to \widetilde{U(1)} \to U(1) \to 1.
\end{equation}
$\widetilde{U(1)}$ is isomorphic to $U(1)$ as a group, but from now on we distinguish it from the original $U(1)$ gauge group of the theory before coupling to $\phi$.
The partition function $Z_{U(1)}[\tilde{X},k]$ is then interpreted as the partition function of the $\widetilde{U(1)}$ gauge theory with the background $B$ of the $\mathbb{Z}_q$ subgroup of the $U(1)$ one-form symmetry \cite{Gaiotto:2014kfa} as $B = k \mathrm{PD}[\tilde{X}]$,
 where $\mathrm{PD}[\tilde{X}]$ is the Poincar\'e dual of the fundamental class $[\tilde{X}]$.
Summing over $k$ effectively does the quotient of the gauge group from $\widetilde{U(1)}$ to $U(1)$. 

Using the modular transformation of the theta function: $\theta(\frac{\tau}{z},-\frac1\tau) = \sqrt{-\mathrm{i}\tau}e^{\frac{\pi}\tau \mathrm{i} z^2}\theta(z,\tau)$, we can rewrite \eqref{eq:ZXk} as
\begin{equation}
    Z_{U(1)}[\tilde{X},k] = \frac{\sqrt{\pi}}{\sqrt{a}}\sum_{\tilde{\jmath}\in \mathbb{Z}} e^{-\frac{\pi^2}{a}(q \tilde{\jmath}-k)^2},
\end{equation}
which is entirely non-perturbative in $a \propto g^2$ for $k\neq 0$.
Therefore, we have explicitly checked that the replica instanton contributions are non-perturbative.

\begin{figure}[t]
    \centering
    \includegraphics[width = .45\linewidth]{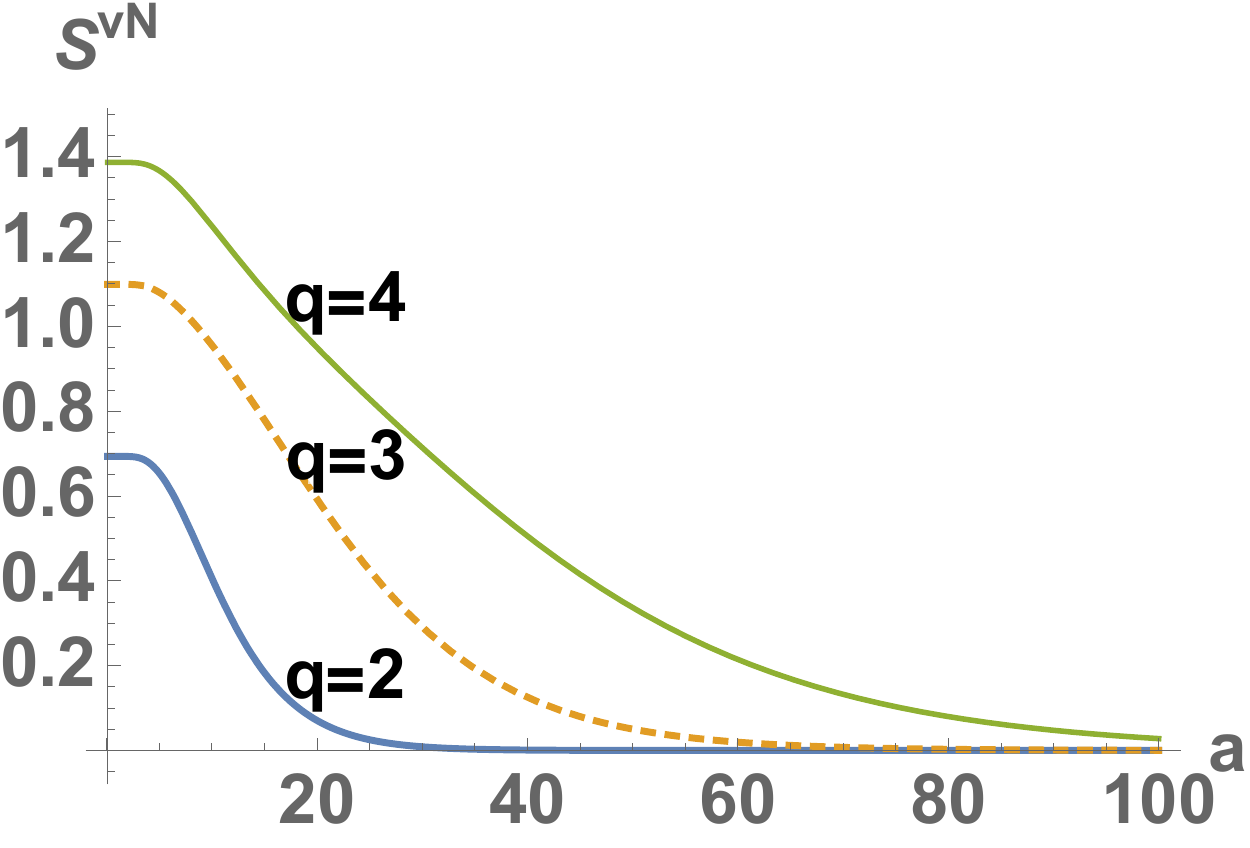}
    \caption{The von Neumann entanglement entropy for $U(1)$ gauge theory coupled to $\mathbb{Z}_q$ scalar reservoir for $q=2,3,$ and $4$. The value at $a=0$ is $\log{q}$ which is the entanglement entropy of the reservoir.}
    \label{fig:SvNplot}
\end{figure}

The entanglement von Neumann entropy $S_\text{vN}$ for $q = 2,3,4$ is plotted in Fig.~\ref{fig:SvNplot}.
In all cases the entropy is $\log q$ when $a=0$ and it becomes zero when $a\to \infty$. 
The value $\log q$ is merely from the reservoir;
the entanglement entropy of an interval (in a circle) in the TQFT of $\mathbb{Z}_q$ valued scalar is $\log q$ as the state $f\sum_{\ell}\ket{\ell}$ is the maximally entangled state among the $q$ states.
As the Euclidean time $t_0$ increases, the state $\ket{X}_\ell$ evolves under the Hamiltonian \eqref{eq:H}.
When $t_0$ is large, the lowest energy eigenstate contained in each $\ket{X}_\ell$ dominates.
Such a state for $\ell\neq 0$ is considered as the state of confining string between the probe charge $\ell$ particle of the $\widetilde{U(1)}$ gauge theory.\cite{Komargodski:2020mxz}
As the pure 2d gauge theory is confined, the confining string has a nonzero tension, and thus the norm of $\ket{X}_\ell$ for $\ell\neq 0$, compared to that of $\ket{X}_0$, vanishes as $a\to \infty$ .
In other words, the confinement of the gauge theory purifies the reduced density matrix $\rho_B$ into the state $\ket{0}_B\bra{0}_B$ in the $a\to \infty$ limit and the entropy gets zero, as shown in the plot.

\begin{figure}[t]
    \centering
    \includegraphics[width = .7\linewidth]{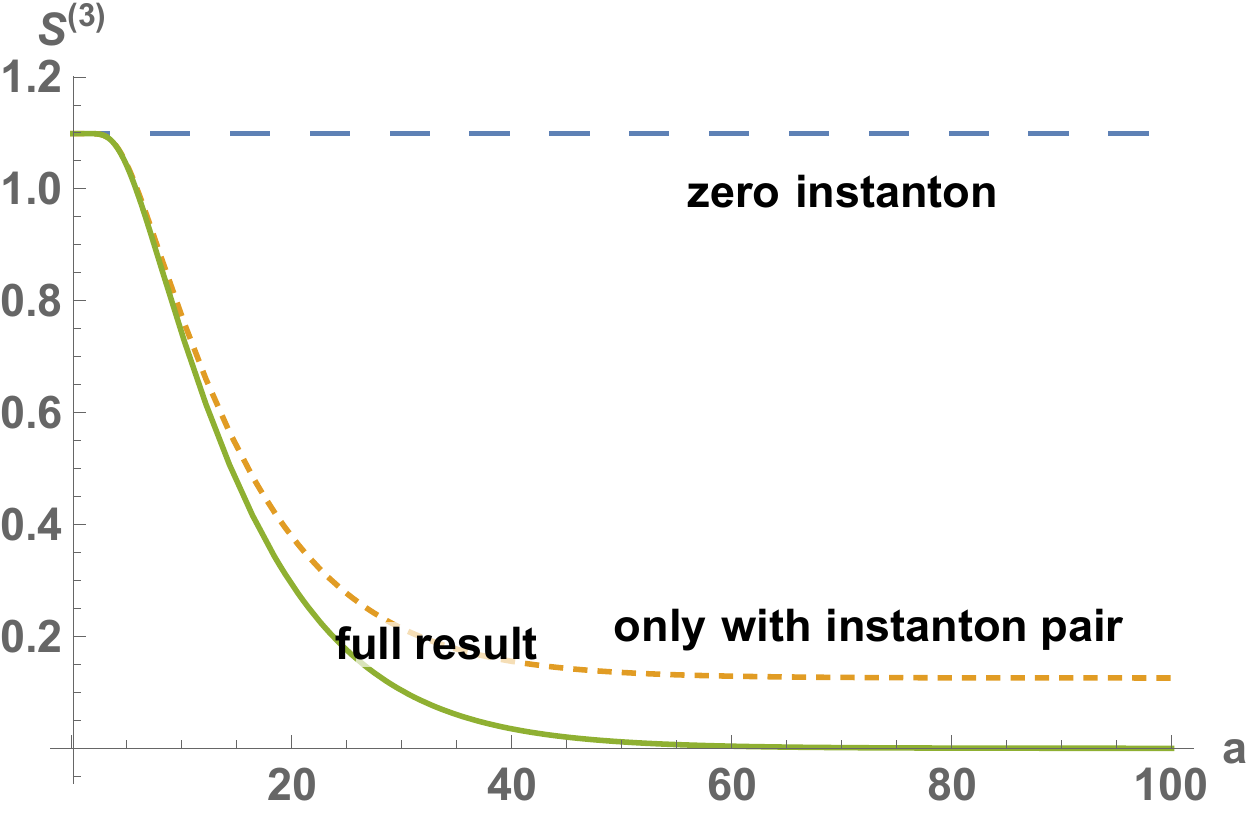}
    \caption{The R\'enyi entanglement entropy $S^{(3)}_B$ for $q=3$. The dotted line does not include the contribution from a triple of instantons, while the dashed line does not include the instantons at all. The solid line is the full result.}
    \label{fig:S3plot}
\end{figure}

To illustrate the effect of replica instantons, we consider the R\'enyi entropy with $n=3$ in the case of $q=3$.
The explicit form of \eqref{eq:ZRn} in this case is
\begin{multline}
    Z[R_3] = 3 (Z_{U(1)}[\tilde{X},0]^3 + 6 \left(\prod_{k=0}^2Z_{U(1)}[\tilde{X},k]\right)\\ + Z_{U(1)}[\tilde{X},1]^3 +Z_{U(1)}[\tilde{X},2]^3).
    \label{eq:ZR3}
\end{multline}
The first term is the zero-instanton contribution, while the next term is from the configuration with a pair of an instanton and an anti-instanton. 
The last two terms are from a triple of instantons or anti-instantons. 
As $q=3$, the total instanton number is constrained only by modulo 3 and thus these configurations contribute.
The R\'enyi entropy $S^{(3)}$ with and without replica instanton contributions is plotted in Fig.\ \ref{fig:S3plot}. 
When all the contributions are included, the entropy approaches zero when $a\to \infty$ as $\rho_B$ becomes pure, while without some of the instanton contributions it does not reach zero.

\begin{figure}[t]
    \centering
    \includegraphics[width = .45\linewidth]{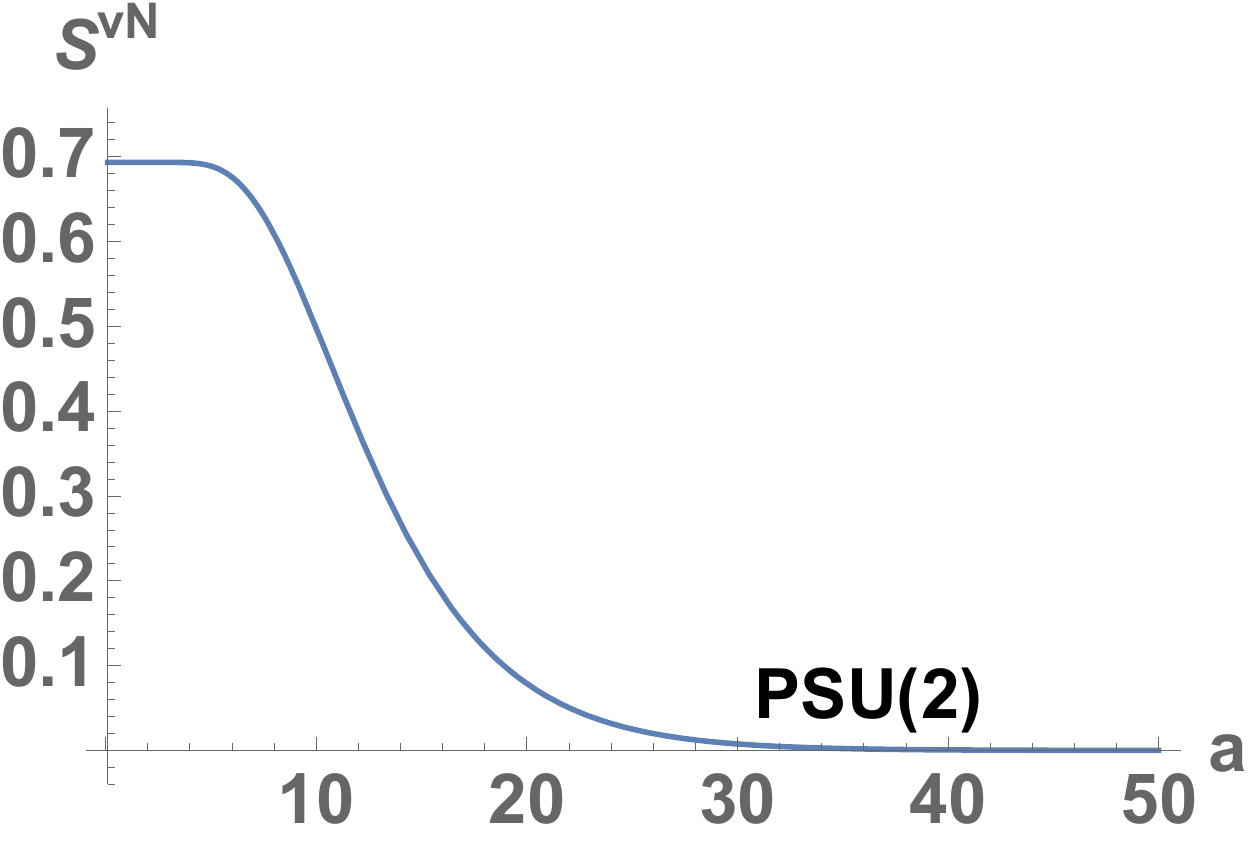}
    \caption{The von Neumann entanglement entropy for $PSU(2)=SO(3)$ gauge theory coupled to $\mathbb{Z}_2$ scalar.}
    \label{fig:SvNPSU2}
\end{figure}

\paragraph{2d pure $PSU(N)$ gauge theory}
The above analysis is easily generalized to a gauge group $G$ with a non-trivial fundamental group.
Here we in particular set $G = PSU(N)$.
In this case, the topological quantity $c(A)$ is taken to be the (generalized) 2nd Stiefel-Whitney class $w_2(A) \in H^2(\tilde{X},\mathbb{Z}_N)$ of the bundle, which is the obstruction for a $PSU(N)$ bundle to be lifted to a $SU(N)$ bundle. As $w_2$ is $\mathbb{Z}_N$ valued, $q$ has to divide $N$ for the coupling \eqref{eq:axion} to be consistent. Here we take $q=N$.
The partition function $Z_{PSU(N)}^{(\ell)}[\tilde{X}]$ is \footnote{
    The Dirichlet boundary condition forces the holonomy along the boundary to be zero. The same effect can be caused by identifying the boundary to a point. Thus the partition function on $\tilde{X}$ is the same as that on a sphere with the same area. 
}
\begin{equation}
    Z_{PSU(N)}^{(\ell)}[\widetilde{X}] = \sum_{R,N(R)=\ell} (\mathrm{dim} R)^2 \, e^{-a C_2(R)},
\end{equation}
where the sum is over the $SU(N)$ irreducible representations $R$ with $N$-ality $N(R) = \ell$, and $C_2(R)$ is the quadratic Casimir.
For the exact results on 2d pure gauge theory, see \cite{Cordes:1994fc,Aminov:2019hwg}.
The contribution from a fixed topological bundle is
\begin{equation}
    Z_{PSU(N)}[\widetilde{X},k] = \frac{1}{N}\sum_{R} (\mathrm{dim} R)^2 \, e^{-a C_2(R)+ 2\pi \mathrm{i}\frac{kN(R)}{N}}.
\end{equation}
This partition function should be understood as the partition function of $SU(N)$ (which is the $\mathbb{Z}_N$-extension of $PSU(N)$) gauge theory with $k$ units of one-form symmetry background.
The entropy $S_B^{\mathrm{vN}}$ is plotted in Fig.~\ref{fig:SvNPSU2}.
As in the $U(1)$ case, the entropy approaches to zero as $a\to \infty$, indicating the confinement of the theory.

\paragraph{Discussion}
In this letter we have discussed the entanglement entropy in a gauge theory coupled with the reservoir of a scalar field by the axion-like coupling \eqref{eq:axion}.
We pointed out that instantons distributed across the replica sheets give non-perturbative corrections to the entanglement entropy, which is analogous to the replica wormholes in gravitational theories.
The explicit computation is done for 2d pure gauge theories coupled with discrete scalars. It is noted that the entropy can be used to detect the confinement in 2d gauge theories. The relationship between the entanglement and confinement is also pointed out in \cite{Klebanov:2007ws} in the holographic context.

It is desirable to do a computation with continuous scalars, where we expect more dynamical phenomena.
Also having a more physically meaningful set up, hopefully with direct analogy to the black hole physics would be interesting.
%Alternatively, one might consider a higher-dimensional gauge theory coupled with a discrete scalar. Such a coupling is considered in \cite{Tanizaki:2019rbk}.
It would also be interesting to consider the large-$N$ limit of the 2d $PSU(N)$ model and its interpretation as a string theory. The entanglement entropy in a large-$N$ 2d Yang-Mills theory is considered in \cite{Donnelly:2019zde}.

Finally, we would like to point out that the "replica instanton" has been considered from 90's, e.g.\ \cite{dotsenko1999griffiths}, in the spin-glass literature. The replica method is used to compute the disordered free energy, and the "instanton" correction represents the "island" of ordered phase appears due to the fluctuation of the coupling. It would be significant to find the connection between this phenomenon and the replica wormholes.

\paragraph{Acknowledgements} KO thanks Zohar Komargodski and Douglas Stanford for useful discussions and comments. 
\bibliography{refs}

\end{document}